\documentclass{mem}
\usepackage{natbib}\usepackage{txfonts}
\usepackage{graphicx}
\usepackage[a4paper]{hyperref}
\idline{75}{282}
\begin{document}
\def\teff{$T\rm_{eff }$}
\def\kms{$\mathrm {km s}^{-1}$}

\title{
Pattern speed evolution and bar reformation
}
 \subtitle{}

\author{
F. \,Combes
          }
  \offprints{F. Combes}

\institute{
Observatoire de Paris, LERMA --
61 Av. de l'Observatoire,
F-75014 Paris, France
\email{francoise.combes@obspm.fr}
}

\authorrunning{Combes}
\titlerunning{Pattern speed evolution and bar reformation}

\abstract{
 Bars in spiral galaxies can weaken through
gas inflow towards the center, and angular momentum
transfer. Several bar episodes can follow one another
in the life of the galaxy, if sufficient gas is accreted
from the intergalactic medium to revive young disks.
  Pattern speeds of the successive bars are different,
due to mass concentration, or increased velocity dispersion
of the remaining stellar component. In the same time, the
spiral galaxy evolves in morphological type. Numerical
simulations are presented, trying to correlate type
and bar pattern speeds.
}
\maketitle{}

\section{Evolution through gas accretion}

In the hierarchical scenario, most of the mass of galaxies is assembled
through mergers of sub-units.  Although this mechanism works
pretty well for early-type galaxies and ellipticals, the interaction between
galaxies is quite  destructive for disks. To account for the high prevalence
of thin disks in today observations, another scenario has to be invoked in parallel,
namely external gas accretion from filaments.  This diffuse, progressive
and smooth mass accretion is much less destructive, and help
to replenish galaxy disks in gas available to new star formation.

Gas accretion can reform cool and thin disks, which are
then unstable to new spiral and bar patterns, in addition to
star formation (Bournaud \& Combes 2002).
The relative role of gas accretion and mergers has 
recently been estimated in numerical simulaitons
(Dekel et al 2008).
The detailed analysis of a cosmological simulation
with gas and star formation 
shows that most of the starbursts are due to smooth flows,
which exceed the merger source by about a factor 10.
Inflow rates are sufficient to assemble galaxy masses
(10-100 M$_\odot$/yr).

These considerable gas flows should have large
consequences on the dynamics of galaxy disks.
In particular, they should accelerate the succession of
patterns in the disk. The role of gas on bars has been
investigated for a long time now (e.g. Friedli \& Benz 1993).

In these early simulations, there was no dark matter component.
The non-axisymmetric patterns produced gravity torques and
the exchange of angular momentum between gas and stars
and from inner to outer parts.
The expulsion of the angular momentum in the outer parts,
allows the gas to flow towards the center. 
This was accompanied by the destruction of the bar
and formation of a triaxial bulge.

\begin{figure*}[t!]
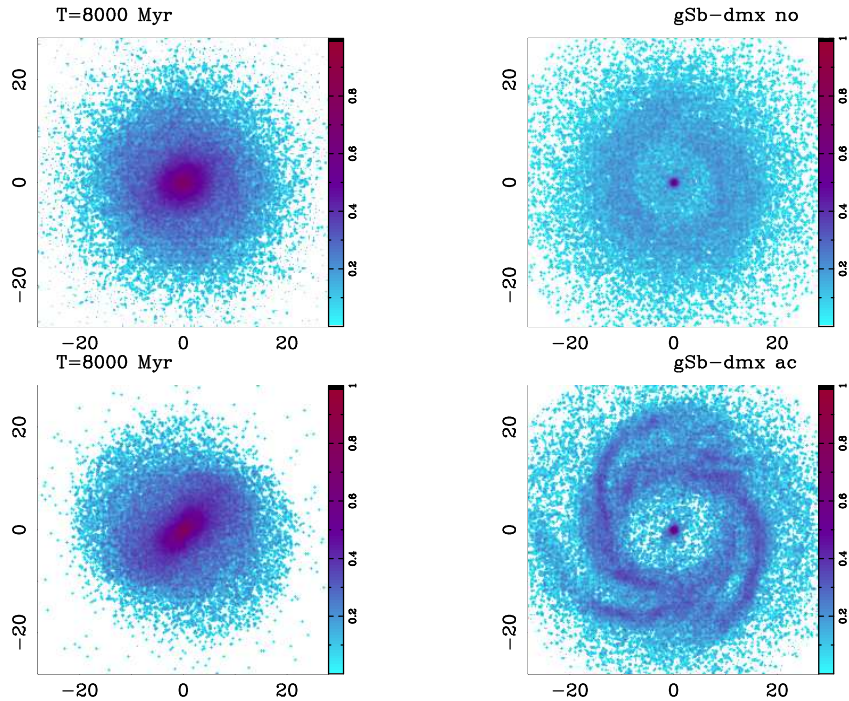

\centering
\includegraphics[clip=true,angle=0,width=11cm]{combes-f1a.ps}
\includegraphics[clip=true,angle=0,width=11cm]{combes-f1b.ps}
\caption{Comparison between gSb galaxy models, without gas accretion
({\bf top}) and with accretion ({\bf bottom}). The gSb model is the maximum
disk model, the gas accretion rate is 5 M$_\odot$/yr, and the snapshot
corresponds to T=8 Gyr. Note that a bar is maintained only in the 
case of gas accretion.
}
\label{fig1}
\end{figure*}

\section{Bar formation and destruction}

The mechanism invoked to account for the observed destruction
of bars due to the gas inflow, was the implied
central mass concentrations (CMC). When a massive
CMC is added like a black hole in the center of a purely
stellar disk, the effect is indeed to
weaken and destroy bars (Norman, Sellwood \& Hasan 1996).
This is attributed to the chaos induced by the superposition 
of CMC and bar on the stellar orbits, as checked by the computation
of surface of sections.

\begin{figure*}[t!]
\centering
\includegraphics[clip=true,angle=-90,width=10cm]{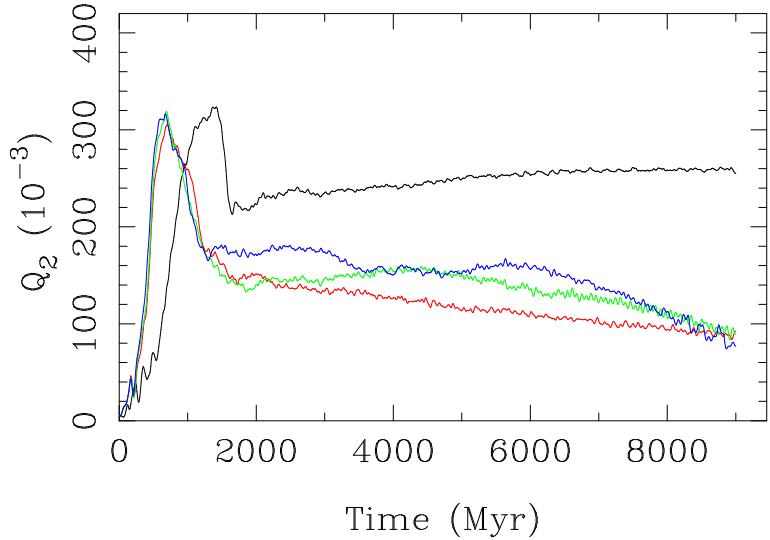}
\caption{Evolution of bar strength, measured as the ratio 
Q$_2$ of the maximal $m=2$ tangential force to the radial force, measured
at a radius of 3.3kpc. The black curve corresponds to the purely stellar run,
the red curve, to the spiral galaxy with initial gas, subject to star formation,
but not replenished. These two curves are respectively the top and the bottom
curves. The other curves corresponds to the models with gas accretion,
5 M$_\odot$/yr for the green one, and 5 M$_\odot$/yr for the blue one.
The galaxy model is an early-type gSa. Note that the presence of
gas triggers bar instability earlier than in the purely stellar run.
}
\label{fig2}
\end{figure*}

\subsection{Bar destruction by gas}

The effect of the gas inflow is not only to produce
a central mass concentration. In fact,
the gas is driven in by the bar torques, but
reciprocally, the gas exerts the opposite torque on the
stars. The angular momentum lost by the gas is taken up 
by the bar wave. Since the latter is a pattern
with negative angular momentum (inside its corotation),
this exchange of momentum  weakens and destroys the bar. 
Simulations show that the gas infall of
 1-2\% of disk mass is enough to transform a bar in a lens
(Friedli 1994, Berentzen et al 1998, Bournaud et al 2005).

\begin{figure}[t!]
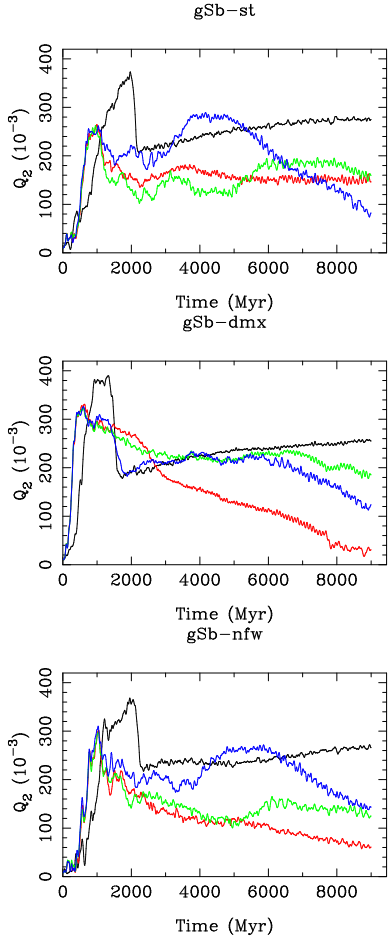

\centering
\includegraphics[clip=true,angle=-90,width=5cm]{combes-f3a.ps}
\includegraphics[clip=true,angle=-90,width=5cm]{combes-f3b.ps}
\includegraphics[clip=true,angle=-90,width=5cm]{combes-f3c.ps}
\caption{Evolution of bar strength, as in Figure 2 (black is the purely stellar run,
red, with gas but no accretion, green with 5 M$_\odot$/yr accretion, and red 10 M$_\odot$/yr),
for the 3 potential models of the gSb galaxy: standard, maximum disk,
and cuspy NFW dark matter halo.
}
\label{fig3}
\end{figure}

\subsection{Reformation of bars}

If it is the gas inflow itself that destroys the bar,
the advantage over the CMC mechanism is that it does
not prevent a new bar to form. As soon as some new gas
is accreted to replenish the disk, and makes it
unstable again to bar formation, there
could be a succession of patterns crossing the disk
(see e.g. Figure 1).  

It works like a self-regulated cycle: the first step being
the formation of a bar in a cold unstable disk, then the
bar produces gas inflow, which 
weakens or destroys the bar. External gas accretion
closes the loop. The gas enters the disk by intermittence:
first it is confined outside OLR until the bar weakens,
 then it can replenish the disk, to make
 it unstable again to bar formation.

This is the mechanism when the dark matter component
may be ignored, in the inner parts of giant galaxies.
When a massive dark matter is taken into account in simulations,
the angular momentum exchange is preferentially done
with the DM particles, due to dynamical friction.
Because of its low mass, the gas plays then 
a minor role in the AM transfer, at least when the
fraction of gas is lower than 8\% (Berentzen et al 2007).

However in DM-dominated simulations, the 
bar is still weakened or destroyed more quickly in the presence of gas.
This is interpreted in terms of the destruction by the gas of
vertical resonances, and with the help of the CMC.
To destroy the bar, the gas fraction must be 
higher in the presence of a massive halo.

\begin{figure}[t!]
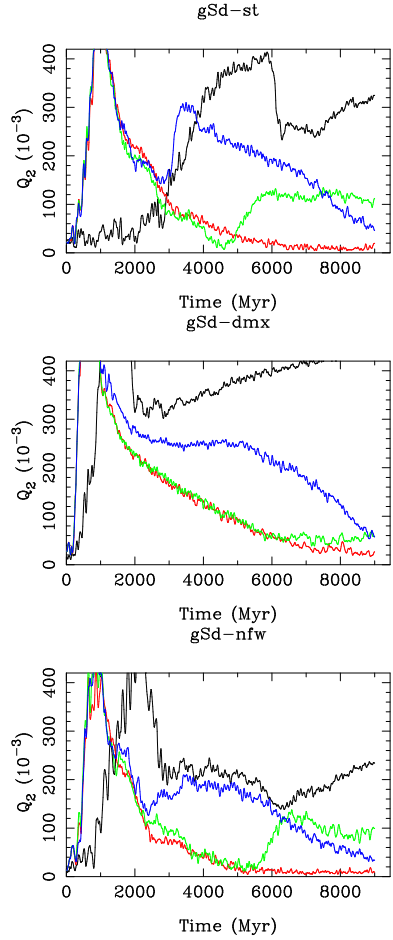

\centering
\includegraphics[clip=true,angle=-90,width=5cm]{combes-f4a.ps}
\includegraphics[clip=true,angle=-90,width=5cm]{combes-f4b.ps}
\includegraphics[clip=true,angle=-90,width=5cm]{combes-f4c.ps}
\caption{ Same as in Figure 3, but for the gSd models.
}
\label{fig4}
\end{figure}

The role of gas is complex since simulations
reveal that it prevents peanut formation.
 The presence of gas implies more chaos in the orbits,
which stop the effect of vertical resonance.
However, the vertical peanut instability is a factor
of bar weakening, in pure stellar disk.
 Since the bar is weakened more quickly in the presence of
gas, the bar weakening mechanism must be different from the
vertical buckling.

\begin{table*}
\caption{Summary of the different models simulated (bulge, disk, halo 
and gas components are respectively described)}
\label{abun}
\begin{center}
\begin{tabular}{lcccccccc}
\hline
\\
Model & M$_b$ & r$_b$ & M$_d$ & r$_d$ & M$_h$ & r$_h$ & M$_{gas}$ & r$_{gas}$ \\
&10$^{10}$ M$_\odot$& kpc & 10$^{10}$ M$_\odot$  & kpc &  10$^{10}$ M$_\odot$&kpc &  10$^{10}$ M$_\odot$& kpc \\
\hline
\\
gSa & 4.6  & 2. &  6.9 & 4. &  11.5 & 10.  & 1.3  & 5. \\
gSb-st & 1.1  & 1. &  4.6 & 5. &  17.2 & 12.  & 0.9  & 6. \\
gSb-dmx & 2.2  & 1. &  9.2 & 5. &  11.4 & 12.  & 0.9  & 6. \\
gSb-nfw & 1.1  & 1. &  4.6 & 5. &  17.2 & 12.  & 0.9  & 6. \\
gSd-st & 0.  & -- &  5.7 & 6. &  17.2 & 15.  & 1.7  & 7. \\
gSd-dmx & 0.  & -- &  11.4 & 6. &  11.4 & 15.  & 1.7  & 7. \\
gSd-nfw & 0.  & -- &  5.7 & 6. &  17.2 & 15.  & 1.7  & 7. \\
\\
\hline
\end{tabular}
\end{center}
\end{table*}

\begin{figure}[h!]
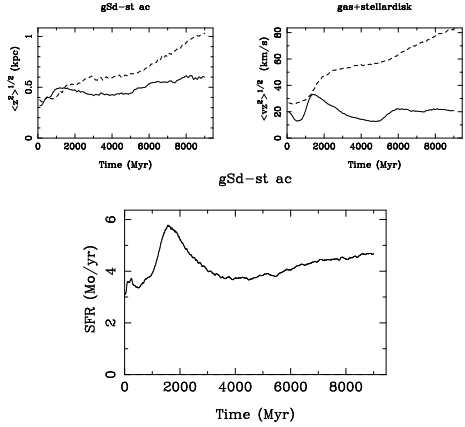

\centering
\includegraphics[clip=true,angle=-90,width=6cm]{combes-f5a.ps}
\includegraphics[clip=true,angle=-90,width=4cm]{combes-f5b.ps}
\caption{Star formation rate versus time ({\bf bottom}), 
average plane thickness ({\bf top left}), and vertical
velocity dispersion ({\bf top right}), for the gSd standard model.
The dash lines correspond to the stellar component, and the full lines to the gas.
}
\label{fig5}
\end{figure}

\begin{figure}[htb!]
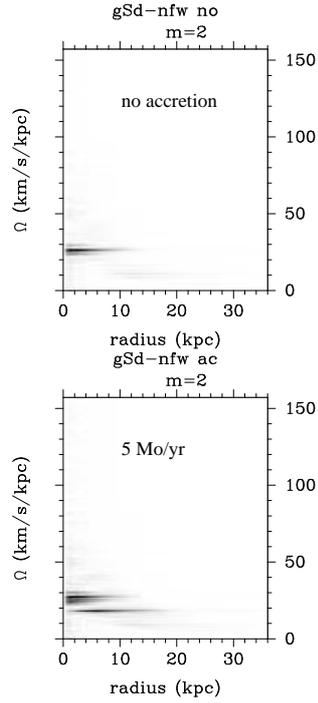

\centering
\includegraphics[clip=true,angle=-90,width=4cm]{combes-f6a.ps}
\includegraphics[clip=true,angle=-90,width=4cm]{combes-f6b.ps}
\caption{ Power spectrum displaying the pattern speed of the 
$m=2$ Fourier component, over the whole simulation of the gSd model,
version NFW potential, without accretion ({\bf top}), 
and with accretion by 5 M$_\odot$/yr ({\bf bottom}).
Without gas accretion, the bar is distroyed rapiddly, and has no time to
slow down, due to dynamical friction. The signature shows a single
pattern speed. For the case with gas accretion, a new bar
is formed, with a smaller pattern speed, and several values
can be read. Note that the bar is more extended, when slower.
}
\label{fig6}
\end{figure}

\section{New simulations of bar reformation}

To further investigate the relative role
of gas and dark matter in the angular momentum 
transfer, and in the bar successive 
formation, weakening and re-formation, we have 
carried out a series of simulations, with different
galaxy models, described in Table 1.
 Three Hubble types are considered, Sa, Sbc and Sd,
and since they are giant spiral galaxies, they
are noted gSa, gSb and gSd.

\subsection{Prescription of the simulations}

The simulations are fully self-consistent, with live dark haloes.
They are using a 3D Particle-Mesh code, based on
FFT, with a useful grid of 
128$^3$ (the algorithm of James (1977) to supress
the Fourier images is used).
The gas is represented by sticky particles,
and a total of 240 000 particles is used.
Similar simulations with TREE-SPH have also been run
(e.g. Di Matteo et al 2007).

The star formation is assumed to follow a Schmidt law, 
with exponent n=1.4, and a density threshold. It is
calibrated for a gas consumption time-scale of 5 Gyr.
We do not assume instantaneous recycling, but
take into account a continuous mass loss, across Gyrs (Jungwiert et al 2001).
The mass loss by stars is distributed through gas on 
neighbouring particles, with a velocity dispersion, 
to schematize the feedback energy.

When external gas accretion is considered,
the gas is deposited at the inner border of the disk, over 4kpc.
At this radius, the gas is assumed to have 
already settled in the plane, with circular velocity.
To cope with large mass variations of the gas component, the number
of gas particules is kept constant, but the individual particle
mass is variable. The same is assumed for the "star" particles,
which can be newly born, or lose mass.

The mass and characteristic radii of all components in
the seven initial conditions, are displayed in Table 1.
The standard model has a disk dominated by dark matter,
but with a core. An NFW model, with a cusp and a central
concentration of c=8  has been run, then a maximum disk model 
(dmx) with equal mass in baryons and DM within the optical
disk. For each galaxy model,
we have 4 runs: a control simulation with pure stellar
disks, without gas, then a run with gas and star formation,
and two other runs with gas accretion, either 5 M$_\odot$/yr,
or 10 M$_\odot$/yr.

The rotation curves are fitted to the observed
typical curves for these Hubble types. For these
giant spiral galaxies, the total mass inside 35kpc 
is 2.4 10$^{11}$ M$_\odot$, with 75\% n DM and 25\% baryons,
for the standard model (and equality between baryons and DM 
for the dmx model).
In the gas accretion runs, the final mass is
17\% or 35\% higher after 9 Gyr.
Due to the varying bulge-to-disk mass ratio
over the sequence, the precession rate $\Omega - \kappa/2$ 
sweeps a large range, so that the expected
pattern speed of the bar patterns are expected
to vary correspondingly.

The bar strength and pattern speed are computed
from Fourier analysis of the potential,
as in Combes \& Sanders (1981):
$$
Q_2= F_\theta(m=2)/F_r 
$$
We take the maximum over the azimuth of $Q_2$, 
estimated at a radius of R=3.3kpc.

\subsection{Angular momentum (AM) transfer}

When the galaxy disk is dominated by the mass
of the dark matter halo, the main 
angular momentum transfer is from baryons to the dark matter
(Athanassoula 2002, 2003). In that sense, the DM halo can favor
(instead of prevent) bar formation.

With light dark matter haloes, the bar is formed
or weakened while AM is exchanged between the inner disk and 
the outer disk or gas.
The exact processus depends on the amount of dark matter
inside the visible disk.

In addition, the pattern speed of the bar decreases
due to dynamical friction against the dark matter particles
(Debattista \& Sellwood 2000).
We observed the same phenomenon with our
standard models, in the case of pure stellar disks.
In the gSa model for instance, the pattern speed $\Omega_b$ 
begins at 50 km/s/kpc, and the bar slows
down until around 10 km/s/kpc in the pure stellar run.
With initial gas, and no accretion, the bar is
quickly weakened, and there is then
three times less angular momentum transfer: 
$\Omega_b$ decreases less, from 50 to 25 km/s/kpc,
in 9 Gyr.

\begin{figure}[htb!]
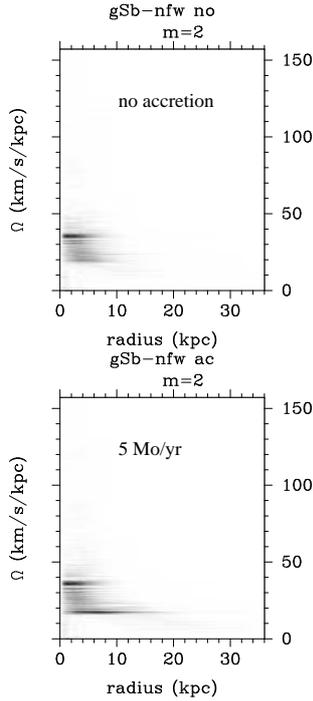

\centering
\includegraphics[clip=true,angle=-90,width=4cm]{combes-f7a.ps}
\includegraphics[clip=true,angle=-90,width=4cm]{combes-f7b.ps}
\caption{ Power spectrum displaying the pattern speed of the 
$m=2$ Fourier component, over the whole simulation of the gSb model,
version NFW potential, without accretion ({\bf top}), 
and with accretion by 5 M$_\odot$/yr ({\bf bottom}).
Note how thick is the spectrum, accumulating 
several bar patterns with slower and slower speeds.
With gas accretion, a new bar
is formed, more extended, with a smaller pattern speed. 
}
\label{fig7}
\end{figure}

\subsection{Comparison of bar strengths}

Figure 2 compares the bar strength of several runs, in the case
of the gSa galaxy model. When the disk is purely stellar,
the bar is long lived (black curve), although
it is suddenly weakened when the peanut resonance
thickens the disk vertically.
When the gas is included (and for this early-tupe
galaxy, it is 5\% of the toal mass), then the bar
is destroyed (red curve). Note that with gas,
the bar instability occurs earlier than in the
pure stellar case, since the disk is then cooler,
and thus more unstable. 
 With gas accretion (5M$_\odot$/yr), the bar 
remains at a higher level, even more with
a double accretion rate (10M$_\odot$/yr),
although the effect remains small. This might
be due to the great stability inferred by the massive
bulge.

The same features occur also in the case of the gSb
model (Figure 3), but now the gas accretion succeeds
to reform a new bar, at the same level as the 
control run, with purely stellar disk.

The case of the gSd model is more complex (Figure 4).
The instability of the disk is large, and dynamical
feedback effects, due to gravitational heating, 
can produce unexpected behaviour.
The dark matter halo stabilises the standard
gSd model; the thin disk, in absence of bulge, 
is vertically unstable, and buckles even before forming 
a bar, which delays by 4 Gyr the bar formation.
In presence of gas, the bar forms in less than 
1 Gyr however. Secondary bar episodes are observed.

Figure 5 shows how gas accretion
can maintain the star formation rate at a
nearly constant value, while it is 
exponentially decreasing without external
gas accretion. The
star formation and corresponding feedback is heating
progressively the stellar disk (visible both in 
vertical height of the disk and vertical velocity dispersion).
The peanut formation heats also the stars,
while the gas can cool down, and retrieve 
an approximately constant z-dispersion.

\section{Pattern speeds}

The normal bar evolution is often
accompanied by a decrease of the pattern speed, and this
phenomenon occurs 
even without dark matter halo and dynamical friction.
This phenomenon was observed in models without DM halo,
or with rigid haloes (Friedli \& Benz 1993).
Stellar orbits are more and more elongated
during bar growth, and their precession rate
decreases (e.g. Combes et al 1990).
This AM exchange with the outer
disk is also accompanied by chaotic escape.
More unstable stellar disks end up with a weaker bar.

Fugure 6 shows the power spectrum of the 
NFW-potential gSd model. The power spectrum is defined 
by the Fourier transform over time of the bar strength,
at each radius:
$$
Power(\omega,r)=\int Q_2(t,r) exp(i\omega t) dt
$$

In the case of no gas accretion, the bar is
quikcly destroyed, and the spectrum
is clear and thin: only one bar at
28 km/s/kpc is seen.
 With gas accretion, a second bar is formed,
but with lower pattern speed, 18 km/s/kpc, and
two features are seen in the power-spectrum.
 In the case of a bar only weakened, but not
destroyed, and slowed
down by dynamical friction, a thick spectrum
is observed, as a tracer of the continuously 
decreasing pattern speed (cf Figure 7).

It is important to note that bars always end up
near their corotation. When the bar is slowed down,
its length increases correspondingly.
Figure 6 shows clearly how the slower bar 
is more extended in radius. It will then be
quite difficult observationnally to retrieve
the history of the bar pattern speed evolution.

As a consequence of AM transfers, mass is
redistributed in the disk, and the surface density
reveals breaks in the radial profiles.
Surface density breaks have been interpreted
as the formation of successive bars by 
Debattista et al (2006). These could 
 be a stellar processus (Pfenniger \& Friedli 1991),
or a gas process, with star formation threshold
(Roskar et al 2008).

\section{Conclusions}

Gas plays a fundamental role in spiral galaxies.
Even when representing only 5-7\% of the total mass inside 35kpc,
the presence of gas weakens or destroys the bar.
It can weaken the peanut, changes the bar pattern speed.

 To account for the bar frequency
observed today, it is essential to take into account gas accretion: 
it provides bar reformation, through 
disk replenishment. The amount of gas accretion required,
is justified and confirmed by cosmological 
simulations.

It is possible that mass assembly could be attributed to smooth accretion
more than mergers: this will solve the problem of
the high frequency of bars observed today, and also the existence of
thin stellar disks, that mergers would over-heat.

Finally, pattern speeds have a tendency to decrease,
first due to the elongation of orbits, and the heating
of the disk, then
 bars are slowed down by dynamical friction against the
dark halo. When bars are reformed, they occur with a lower pattern speed.
This may constrain the amount of dark matter present within the optical disk.
 
\begin{acknowledgements}
I am grateful to E. Corsini and  V. Debattista
to invite me at such an exciting and friendly conference. 
\end{acknowledgements}

\bibliographystyle{aa}

\begin{thebibliography}{}
\bibitem[]{Lia02} Athanassoula, E.: 2002, ApJ 569, L83
\bibitem[]{Lia03} Athanassoula, E.: 2003, MNRAS 341, 1179
\bibitem[]{Berentzen98} Berentzen I., Heller, C. H., Shlosman, I., Fricke, K. J.: 1998, MNRAS 300, 49
\bibitem[]{Berentzen07} Berentzen I., Shlosman, I., Martinez-Valpuesta, I., Heller, C.: 2007, ApJ 666, 189
\bibitem[]{} Bournaud F., Combes F.: 2002, A\&A 392, 83
\bibitem[]{B05} Bournaud F., Combes F., Semelin B.: 2005, MNRAS 364, L18
\bibitem[]{Combes81} Combes F., Sanders R.H.: 1981, A\&A  96, 164
\bibitem[]{Combes90} Combes F., Debbasch F., Friedli D., Pfenniger D.: 1990, A\&A  233, 82
\bibitem[]{Victor00} Debattista V.P., Sellwood J.: 2000, ApJ 543, 704
\bibitem[]{Victor06} Debattista V.P., Mayer, L., Carollo, C. M. et al.: 2006, ApJ 645, 209
\bibitem[{Dekel et al. (2008)}]{dekel08} Dekel, A., Birnboim, Y., Engel, G. et al. 2008,  sub, astro-ph/0808.0553
\bibitem[]{}Di Matteo P., Combes F., Melchior A-L., Semelin B.: 2007: A\&A 468, 61
\bibitem[]{Friedli93} Friedli, D., Benz W.: 1993, A\&A 268, 65  
\bibitem[]{Friedli94} Friedli, D., 1994, in Mass-Transfer Induced Activity in Galaxies, Ed I. Shlosman,
    Cambridge University Press,  p.268
\bibitem[]{} James, R.A. 1977: J. Comput. Phys. 25, 71
\bibitem[]{} Jungwiert B., Combes F., Palous J.: 2001, A\&A 376, 85
\bibitem[]{Norman96} Norman C., Sellwood J.A., Hasan H.: 1996 ApJ 462, 114
\bibitem[]{Daniel01} Pfenniger, D., Friedli, D.: 1991, A\&A, 252, 75
\bibitem[]{Roskar08} Roskar R., Debattista V.P., Stinson G.S. et al: 2008, ApJ 675, L65

\end{thebibliography}

\end{document}